\documentclass{sig-alternate-05-2015}

\begin{document}

% Copyright
\setcopyright{acmcopyright}
%\setcopyright{acmlicensed}
%\setcopyright{rightsretained}
%\setcopyright{usgov}
%\setcopyright{usgovmixed}
%\setcopyright{cagov}
%\setcopyright{cagovmixed}

% DOI
\doi{}

% ISBN
\isbn{}

%Conference
\conferenceinfo{WebSci '17}{June 26--28, 2017, Troy, NY, USA}

%\acmPrice{\$15.00}

%
% --- Author Metadata here ---
\conferenceinfo{WebSci}{'17 Troy, NY, USA}
%\CopyrightYear{2007} % Allows default copyright year (20XX) to be over-ridden - IF NEED BE.
%\crdata{0-12345-67-8/90/01}  % Allows default copyright data (0-89791-88-6/97/05) to be over-ridden - IF NEED BE.
% --- End of Author Metadata ---

\title{The Fake News Spreading Plague: Was it Preventable?}

\numberofauthors{2} %  
\author{
\alignauthor Eni Mustafaraj\\
       \affaddr{Department of Computer Science}\\
       \affaddr{Wellesley College}\\
       \affaddr{Wellesley, MA, USA}\\
       \email{emustafaraj@wellesley.edu}  
% 2nd. author
\alignauthor Panagiotis Takis Metaxas\\
       \affaddr{Department of Computer Science}\\
       \affaddr{Wellesley College}\\
       \affaddr{Wellesley, MA, USA}\\
       \email{pmetaxas@wellesley.edu}
}

\maketitle
\begin{abstract}
In 2010, a paper entitled ``From Obscurity to Prominence in Minutes: Political Speech and Real-time search'' \cite{mustafaraj:websci10} won the Best Paper Prize of the Web Science 2010 Conference. Among its findings were the discovery and documentation of what was termed a ``Twitter-bomb'', an organized effort to spread misinformation about the democratic candidate Martha Coakley through anonymous Twitter accounts. In this paper, after summarizing the details of that event, we outline the recipe of how social networks are used to spread misinformation. One of the most important steps in such a recipe is the ``infiltration'' of a community of users who are already engaged in conversations about a topic, to use them as organic spreaders of misinformation in their extended subnetworks. Then, we take this misinformation spreading recipe and indicate how it was successfully used to spread fake news during the 2016 U.S. Presidential Election. The main differences between the scenarios are the use of Facebook instead of Twitter, and the respective motivations (in 2010: political influence; in 2016: financial benefit through online advertising). After situating these events in the broader context of exploiting the Web, we seize this opportunity to address limitations of the reach of research findings and to start a conversation about how communities of researchers can increase their impact on real-world societal issues. 
\end{abstract}

% add ACM categories here at some point

%\printccsdesc


\section{Introduction}
\subsection{The Anatomy of a political Twitter-Bomb}
It was January 15, 2010, when a group of nine Twitter accounts, all created within an interval of 15 minutes, with names such as  ``CoakleyAgainstU'', ``CoakleyCatholic'', etc. These were accounts with names starting with the name of the democratic candidate Martha Coakley who was running in the Special Election for the Massachusetts U.S. Senate seat. The newly minted accounts sent 929 tweets addressed to 573 unique users in the course of 138 minutes. All the tweets contained a URL to the same website {\tt http://coakleysaidit.com}, % do not make it clickable
(also registered on January 15, 2010),
that showed video and audio from a speech by Martha Coakley, taken out of context, to advance the false claim that she is against the employment of Catholics in the emergency room.  

\begin{figure}
\centering
\includegraphics[width=3in]{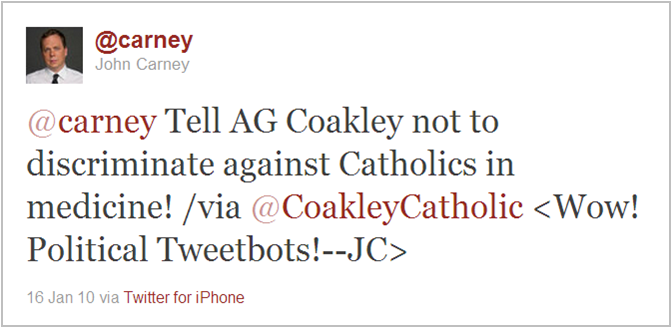}
\caption{The journalist John Carney (at that time with CNBC), received one of these ``reply-tweets'', which he retweeted adding a comment expressing his surprise, ``<Wow! Political Tweetbots!--JC>'', because this was an unknown phenomenon at that time on Twitter. Carney deleted the URL of the original tweet.}
\label{fig:carney}
\vskip -8pt
\end{figure}

The nine accounts were sending a tweet per minute and repeating more or less the same content – reasons to be flagged as a spamming account. Twitter discovered the automated tweets  and consequently suspended all nine accounts. Their existence and their misinformation attack would have gone unnoticed had it not been for one fortunate circumstance: we were collecting all tweets containing the names “coakley” and “brown,"  for Martha Coakley and Scott Brown respectively who were the two candidates for the senate election, in real-time during the week leading to the election.
 But, the tweets sent by these anonymous accounts were not simple tweets, they were so-called ``reply tweets'', tweets directed to particular users. Why? Because a new account on Twitter doesn't have any followers. Tweets sent by such an account will not be read by anyone. Thus, addressing the tweets to a particular user makes it likely that the tweet will be read. But to what user do you send the tweet, out of the millions that are on Twitter? This is when a common spamming technique on Twitter comes in handy: reply to users who used certain desired keywords in their tweets, that is, to users already attuned to the topic. Our analysis of the recipients of these ``reply tweets'' revealed that 96\% of them had been tweeting about the MA senate race in the four-hour interval between the time the anonymous accounts were created and when they started to send the ``reply tweets''. Almost 25\% of the users who received a tweet (143 out of 573) retweeted the message. A screenshot from one of the retweets is shown in Figure~\ref{fig:carney}. We chose to show this tweet, because the user is a well known journalist\footnote{John Carney, \url{https://twitter.com/carney}.} and experienced  user who joined Twitter on March 2007. His surprise at the message indicates the novelty of this technique at the time. The retweets had the effect that the followers of the retweeters were likely exposed to the misinformation, which they would have not seen otherwise. This is because the messages didn't include hashtags, a common way to group together tweets about a topic. Our estimation of the audience size, based on the followers of the retweeters, amounted to 61,732 Twitter users.    

\subsection{A recipe for spreading misinformation on Twitter}
All the facts presented in the previous subsection were part of \cite{mustafaraj:websci10}. What we didn't do in that paper was to summarize our findings in an easily memorable recipe, which contains the steps used by the propagandists in spreading their misinformation on Twitter. We're providing this recipe for the first time in this paper.

\begin{table}[htb]
\centering

\begin{tabular}{|l|p{6cm}|}
\hline
\textbf{Step 1} & Register a domain name for a new website, for example: {\tt http://coakleysaidit.com} \\ \hline
\textbf{Step 2} & Create anonymous accounts, for example: \textit{CoakleySaidWhat}, etc. \\ \hline
\textbf{Step 3} & Identify a community of users interested in the topic, for example, the MA Senate Election race.                                      \\ \hline
\textbf{Step 4} & Target members of this community with messages, for example, reply to users providing link to website.\\ \hline
\textbf{Step 5} & Wait for members of community to spread message via retweets in their organic subnetworks. \\ \hline
\end{tabular}
\caption{A recipe for spreading misinformation on Twitter via a ``Twitter-Bomb''.}
\label{tab:twitterbomb}
\end{table}

%\textbf{Note to self and Takis:} Should we say something here about how this research inspired Truthy.

Our discovery attracted the attention of both journalists and other researchers. A team at Indiana University, headed by Fil Menczer, developed Truthy \footnote{Truthy, now known as OSoMe,  \url{http://truthy.indiana.edu}}, a system that collects Twitter data to analyze discourse in near real-time \cite{truthy}. In addition, our team at Wellesley developed Twitter Trails \footnote{TwitterTrails.com, \url{http://twittertrails.com}}, a system that can be used to monitor the spreading of rumors on Twitter \cite{TwitterTrails}. This focus on Twitter is justified by the fact that it provides APIs for researchers to collect and analyze its data, as well as the public nature of conversations on Twitter. Both these features are missing on Facebook (not entirely, but they are severely limited), thus, only Facebook employees are able to study them. As evidence, see \cite{rumor-cascades}. Meanwhile, researchers not affiliated with the company have almost no opportunities to study information spreading on Facebook, especially that of rumors, hoaxes, and recently fake news, a topic to which we turn our focus now. 

The term ``fake news'' refers to lies presented as news, that is, falsehoods online formatted and circulated in such a way that a reader might mistake them for legitimate news articles. ``Fake news'' has been around since ancient times, but information technology has made it possible to produce and consume it on a massive scale. Such articles appear on a variety of little known websites, then turn a profit by competing for clicks as advertisements on social media sites. In order to be successful in attracting user attention, they present a fake story of political nature, religious nature, or anything with an emotional appeal. Typically, fake news stories, which may or may not have some remote connection to reality, are planted on social media sites using provocative titles and images. ``Click bait'' attracts the attention of unsuspecting social media users who click on links to these stories thinking they are visiting a legitimate news site. These engaged users are drawn in by the emotional appeal, while fake news providers get a share of advertising money from each click.

\subsection{Spreading Fake News on Facebook}
After the surprise results of the 2016 U.S. Presidential Election, the American media directed its ire at Facebook and Google, as in this NY Times piece \cite{nytimes:2016} written by the Editorial Board, on November 19, 2016:

\begin{quote}
Most of the fake news stories are produced by scammers looking to make a quick buck. The vast majority of them take far-right positions. But a big part of the responsibility for this scourge rests with internet companies like Facebook and Google, which have made it possible for fake news to be shared nearly instantly with millions of users and have been slow to block it from their sites.
\end{quote}

This criticism is only partly correct. Facebook had been working toward fixing (or containing) the spread of hoaxes on the site at least since January 2015, almost two years before the election \cite{fb:2015}. They defined a hoax as a form of News Feed spam report that includes scams (``Click here to win a lifetime supply of coffee''), or deliberately false or misleading news stories (``Man sees dinosaur on hike in Utah''). As we can notice from this definition, in 2015, the phrase ``fake news'' wasn't being applied yet to the kind of false stories that flooded Facebook in the weeks before the election.

However, since it was difficult for independent researchers to know the extent to which Facebook users were affected by this issue, everything continued more or less as before, and Facebook was alone in its fight. This changed in early 2016, when the online publication BuzzFeed took an interest on Facebook's unsuccessful efforts to deal with the problem. In an article published in April 2016, BuzzFeed proclaims: ``it is the golden age of fake news'' \cite{BF-a1}. The article reveals that BuzzFeed had conducted a study of fake news that was spreading via nine known fake news sites such as the \textit{National Report}, \textit{Huzlers}, or \textit{Empire News}, using the services of the company \textit{Crowdtangle}, specialized in measuring social engagement. The findings revealed that while traffic for these sites had gone down for a while during 2015, in early 2016, it had started picking up again. The article also interviewed Allen Montgomery -- a fake identity for Jestin Coler, the creator of a factory of fake news websites, as \cite{npr:16} revealed after the election. Coler's interview reveals some of the tricks of the trade of fake news, and points out why he believes he can win over Facebook:

\begin{quotation}
Coler believes Facebook is fighting a losing battle, doomed to fail no matter what it does. ``They can shut down \textit{National Report}. Of course they can do that,'' he said. ``But I could have 100 domains set up in a week, and are they going to stop every one of those? Are they now going to read content from every site and determine which ones are true and which ones are selling you a lie? I don't see that happening. That's not the way that the Internet works.'' \cite{npr:16} 
\end{quotation}

\begin{figure}
\centering
\includegraphics[width=\columnwidth]{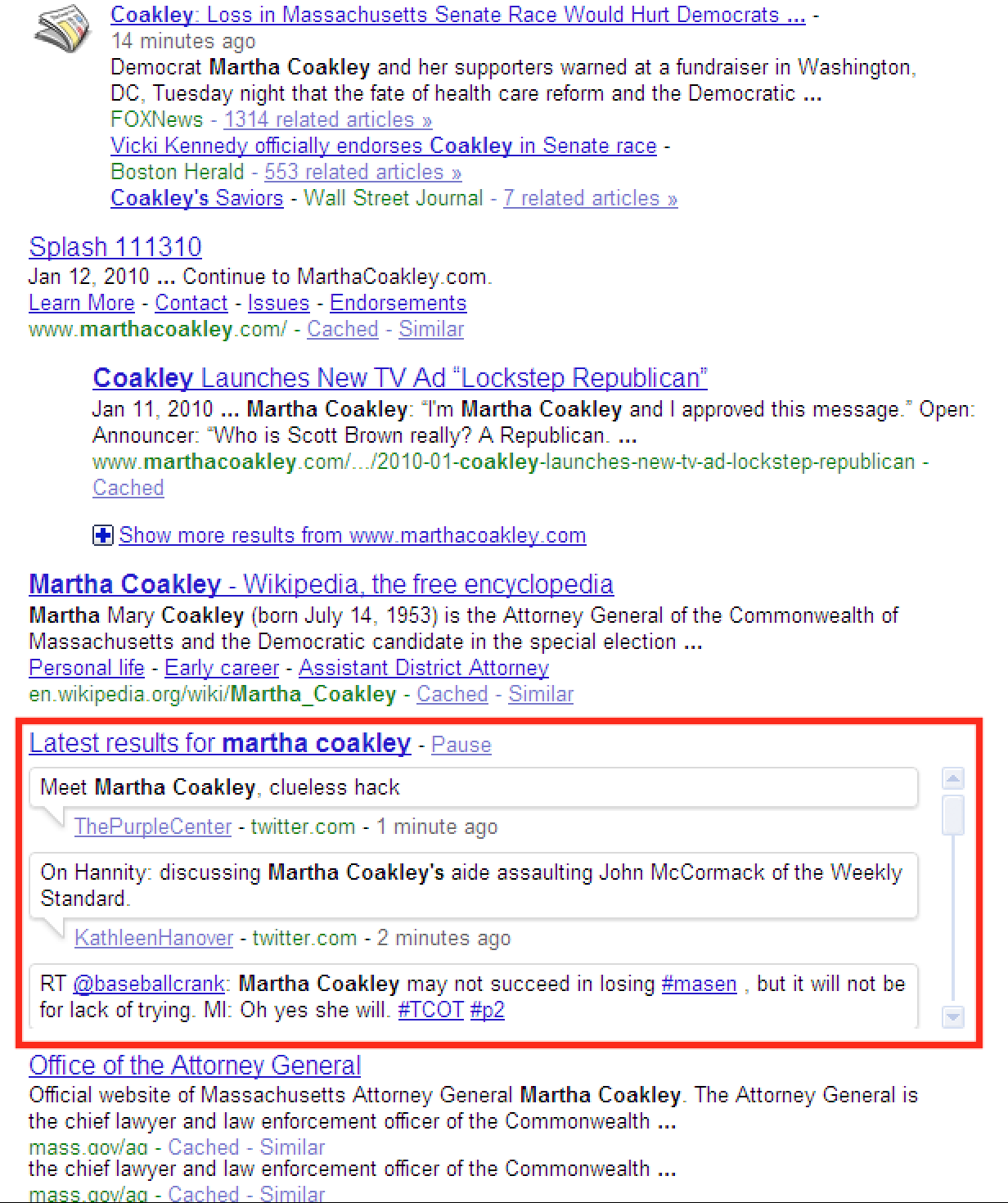}
\caption{Screenshot from Google search results about Martha Coakley on Jan 12, 2010. Notice in highlighted red, the tweets attacking Coakley. This was a finding from \cite{mustafaraj:websci10} on how Google was inadvertently giving premium space to political propagandists, in an effort to have ``fresh'' and relevant search results.}
\label{fig:coakley}
\end{figure}

Despite this sounding of alarm bells by BuzzFeed (as early as April 2016), things got only worse with fake news on Facebook. We counted at least 25 articles published on the topic of fake news from April to November 2016 on BuzzFeed, culminating with the story of ``How teens in the Balkans are duping Trump supporters with fake news'', published on November 3, 2016 and followed up by the related piece on ``How Macedonian spammers are using Facebook groups to feed you fake news''. These two articles provide details about one of the fake news factories operated by young people in the small town of Ceres, Macedonia, that targeted Facebook users in the United States. After reading these news articles (and others on the same topic), we noticed the clear similarities to the process that lead to the Twitter-bomb against Martha Coakley in 2010. In fact, we are able to map the steps in the two recipes one to one, as shown in Table~\ref{tab:facebook}. 

\begin{table}[htb]
\centering
\begin{tabular}{|l|p{6cm}|}
\hline
\textbf{Step 1} & Register web domains for lots of related websites, with catchy names such as: {\tt http://TrumpVision365.com}, see \cite{craig:11-16}. \\ \hline
\textbf{Step 2} & Create Facebook accounts of fictitious people, e.g, Elena Nikolov or Antonio Markoski, see \cite{craig:11-08-16}. \\ \hline
\textbf{Step 3} & Identify and join a Facebook group about a political candidate, e.g., ``Hispanics for Trump'' or ``San Diego Bernicrats'', see \cite{craig:11-08-16}.                                      \\ \hline
\textbf{Step 4} & Target members of the Facebook group with posts, by linking to the fake news website stories, see \cite{craig:11-08-16}.\\ \hline
\textbf{Step 5} & Wait for members of the group to spread the fake news in their organic subnetworks, by sharing and liking it. \\ \hline
\end{tabular}
\caption{The recipe for spreading ``fake news'' on Facebook in the wake of the 2016 U.S. Presidential election. It contains the same steps as the recipe shown in Table~\ref{tab:twitterbomb}. }
\label{tab:facebook}
\end{table}

This similarity should not be surprising. Once a spamming technique has been proven successful, it is easily replicated, since the knowledge about its working is also shared on the Internet. What should surprise and worry us is the fact that researchers and web platform developers may also know about such techniques, but they do little to warn and educate the public of the consequences. It is also unfortunate that tech companies who have been used to facilitate this practice, do not act pro-actively or effectively in stopping it. As an example of ineffective action we refer to the way Facebook handled the accusation that its news verification was not balanced. We discuss it in a next section.

\begin{figure}
\centering
\includegraphics[width=\columnwidth]{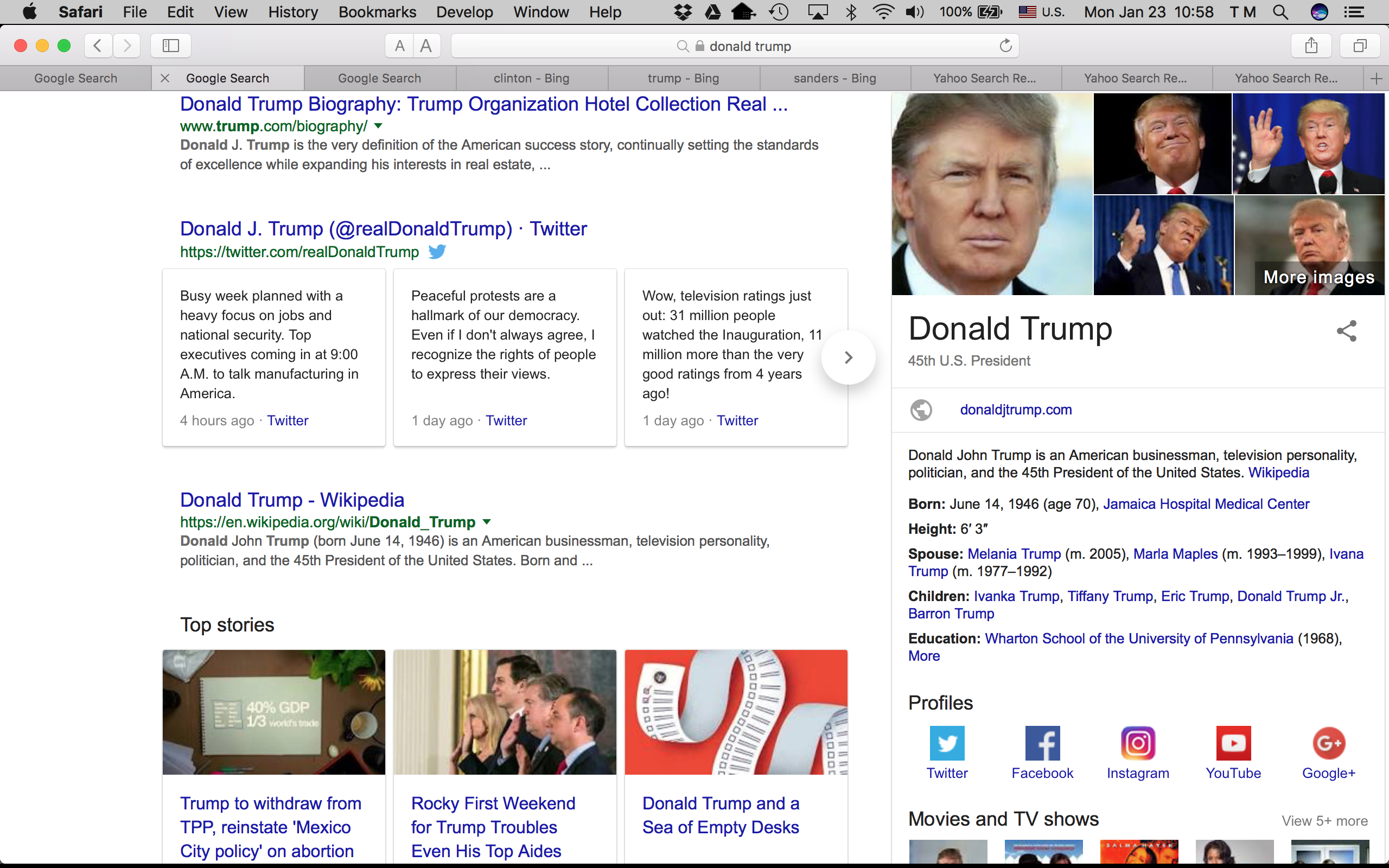}
\caption{Screenshot from Google search results about Donald Trump on Jan 23, 2017. In addition to the many sections to the page (such as the ``featured snippet'' on the right column), notice how the tweets shown above the fold belong to Trump himself.}
\label{fig:trumpGoogle}
\end{figure}

\pagebreak
\section{From Propaganda to Fake News}

We should not give the reader the impression that online propaganda started with Twitter-bombs or Facebook fake news. In fact, it is much older than Social Media, it is as old as the Web itself. However, before the development of search engines it was not easy {\em for propaganda to discover you}. 
Search engines made it easy for propagandists to spread their message using techniques we now call {\em Web Spam} \cite{SpamPropaganda}. Advertisers, political activists and religious zealots of all kinds have been busy modifying the structure of the Web in an effort to promote their own biased results over the organic, unbiased results. The Search Engine Optimization (SEO) industry has grown out of this effort, sometimes using unethical techniques to promote their message, and search engines have been continuously evolving to fend off these attacks.

In much of the first decade of the new millennium, search engines tried to defend against Web Spam, but the spammers were successful in circumventing their defenses by using first ``Link Farms'' and later ``Mutual Admiration Societies'', collections of web sites that would intentionally link to each other in order to increase everyone's PageRank \cite{SpamPropaganda}. Even when Google was reportedly using up to 200 different signals to measure quality \cite{Google200features}, professional SEOs would manage to get sites like JC Penney's at the top of search results \cite{nytimes:JCPenney}. Google's ingenious solution to the problem of ``unfair competition for high placement on a page'' was the introduction of the advertising model of AdWords and AdSense that gave spammers an opportunity to make money while following rules. That seemed to work for a while. However, these financial incentives were so lucrative that they provided a reason for {\em anyone} to have a presence on the Web, especially if that Web presence managed to attract clicks and thus, ad dollars. This led to ``click bait'' and to the creation of ads masquerading as outrageous (fake) news stories, as we discussed in the previous section.

But as Search Engines and Social Media evolve, so do the propagandistic techniques. The rise of three recent methods of spamming in the wake of the U.S. presidential elections, ``fake news'', ``featured snippets'' manipulation \cite{outline:2017}, and ``auto-completion revelations'' \cite{nyt:2012} are the latest chapters in spreading propaganda through search engines and social media so that it will find you. As a community of researchers we need to embrace the challenge of documenting and understanding these phenomena as well as finding ways to make these issues known to the platforms providers. Journalists and non-specialists also need to be informed, as they sometimes give credence to conspiracy theories, confusing the public \cite{wapo:epstein}.

\section{Research that Informs Design}
It is important for researchers, journalists and web users to continue to pay attention to the information and misinformation they encounter on the Web, be it on Google, Twitter, or Facebook. In this section, we discuss how research results and their publicizing lead over time to changes in the design features of these systems, addressing the exhibited weaknesses.

\subsection{The Evolution of Real-Time Search Results}
The central finding that gave the title to \cite{mustafaraj:websci10} was the manipulation of Google real-time search results through repetition of Twitter posts by accounts (real users or bots) supporting a particular candidate. In December 2009 (just one month before the MA special election for the U.S. Senate seat in 2010), Google followed Bing in introducing ``real-time search results'', placing social media messages in one of the top positions search results for a relevant search query. These messages came mostly from Twitter, which had an easy-to-use API to pull the tweets. Tweets appearing in the search results were those that had been recently posted. That created the opportunity for political propagandists to exploit the search rankings creating a \textit{Twitter-enabled Google Bomb}. As we documented in our paper, the manipulators were repeating the same messages, something also allowed by Twitter, over and over to increase the volume of conversation about a particular topic and keep it fresh for search engines to include in their real-time results. Repetition of a message would be annoying to the propagandist followers, but the target was not their followers' feed; it was Google and Bing, and through them, the public. 

We can see these highly-placed tweet messages from random Twitter accounts in a screen shot for Martha Coakley's search results, taken in January 2010 (Figure~\ref{fig:coakley}). During 2010, Google eventually recognized that giving anonymous social media accounts a premium spot in its search results was not in line with its goals for reliable information and for a few years this feature disappeared. However, it has come back again, but in a different format: when searching for a person, it will pull up tweets from their timeline, as opposed to tweets about them, as exemplified in Figure~\ref{fig:trumpGoogle}. This is a great improvement, because it prevents actors who have an interest to promote their adversarial messages about an individual or product to receive an unearned spot at the top of the search results.

%A section on Google's use of the knowledge graph could go here.

\subsection{The Evolution of Retweeting}
In \cite{mustafaraj:websci10}, we included the following observation at the end of Section 4:
\begin{quotation}
Our experiments with Google real-time search have shown
that, even though Google doesn't display tweets from users
that have a spammer signature, it does display tweets from
non-suspected users, even when these are retweets coming
from spammers. Thus, simply suspending spamming attacks
is not sufficient. There should be some mechanism that
allows for retroactively deleting retweets of spam and some
mechanism that labels some Twitter users as enablers of
spam.
\end{quotation}

At that time (in 2010), Twitter didn't have an easy way to quote a tweet and it allowed users to edit the original tweet text when retweeting, as the tweet shown in Figure~\ref{fig:carney} indicated. That design feature turned out to be very problematic, among others for the reason mentioned in the quote above: deleted spam tweets lived in the retweets of other Twitter users. However, it was also problematic because users often were purposefully changing the meaning of the text they were retweeting \cite{mustafaraj:aaai11}. Most of this re-purposing was possible via third-party applications that were very popular in the early years of Twitter. These applications were shut down over the years and nowadays Twitter doesn't allow the editing of a tweet that is being retweeted. Additionally, if the original is deleted, the retweet is also deleted, while in a quoted retweet, the text ``This tweet is unavailable.'' will show in place of the deleted tweet.

% \textbf{Note to self and Takis:} Some concluding observations about this, especially mentioning how the fake news plague has forced Facebook to make changes, such as the use of "Disputed" label for fake news stories. 
% Screenshot: \ref{facebook_marks_fake_news}

\begin{figure}
\centering
\includegraphics[width=\columnwidth]{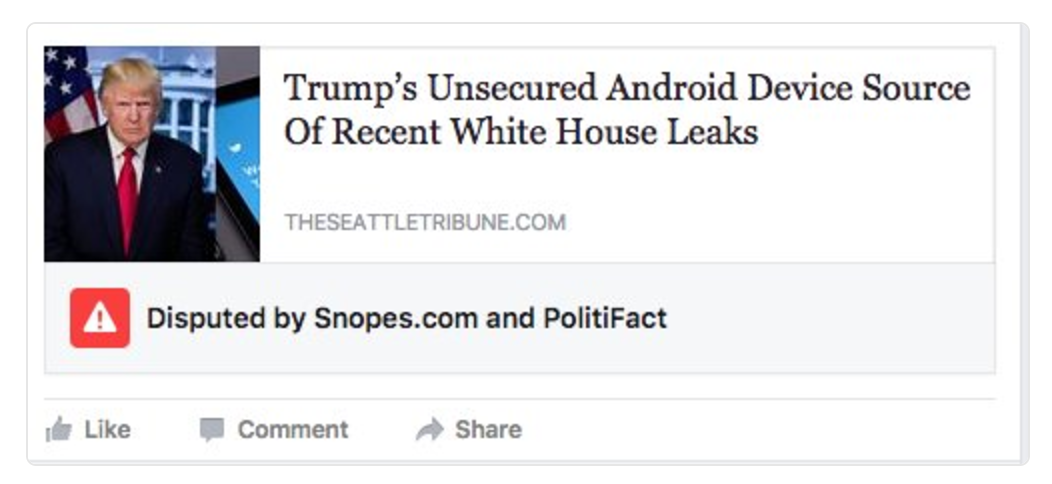}
\caption{Facebook recently moved into implementing a system of warning towards sharing news items that have been disputed.}
\label{fig:facebook-disputed}
\end{figure}

% \subsection{From Propaganda to Fake News and more}
% \textbf{The point of this section is to address what we say in the title: "no need to be surprised", these tactics have always been there, what has changed is the delivery method and the reach they have due to the one single platform for 1 billion people. We also need to talk about the "Rumor Cascades" paper by Adamic \cite{rumor-cascades}. }

% %Any other paper that the WebSci community might know? I found one from WebSci'16, written by Microsoft (not Google) employees: "Information Dissemination in Heterogeneous-Intent Networks" \cite{das:WebSci14} -- Takis, do you have time to skim it? -- Eni, I did look at it but did not see something relevant.}

\subsection{The Evolution of Fake News on Facebook}
% But while our community efforts have been focusing on Twitter and Google, Facebook had a much greater misinformation influence on cyberspace. 
% Major problem for studying facebook's algorithms is the fact that they do not provide an effective way for researchers to collect and analyze their data. The only studies that we are familiar with are those that have been released through research by facebook researchers. 

The proliferation of fake news on Facebook achieved new levels once Facebook made a big change in how its algorithm for the Trending News feature worked. Before August 2016, when this change took effect, the Trending News feature was being curated by human editors, who filtered out unreliable sources and chose neutral or balanced sources for breaking news stories. However, when the online tech blog Gizmodo posted an article in May 2016 \cite{gizmodo:16}, in which former employees of the Trending News lamented anti-conservative bias, Facebook reacted hastily and clumsily.  Worried about potential lawsuits for suppressing freedom of speech, it fired its team of human editors and replaced them with Machine Learning algorithms. Given their prior research on rumor cascades on facebook \cite{rumor-cascades} one would expect that technical experts had a better understanding on the limitations of AI algorithms, but maybe they were not involved in the decision \cite{buzzfeed:renegadeFBemployees}. It didn't take long after that change for fake news to start achieving Trending News status, as BuzzFeed reported on August 30, 2016 \cite{craig:08-16}. Despite BuzzFeed's relentless reporting on the fake news plague throughout the pre-election season, the rest of the media and the public didn't tune in into this conversation until after the election. 

Facebook initially disputed that it had a fake news problem, claiming that it accounts for only 1\% of the news stories. However, the company changed course under the increased and sustained public pressure and introduced new features in its interface and the algorithms to address the issue \cite{fb:2016}.

One important feature that has rolled out recently is the labeling of news posts as ``Disputed'' via fact-checking, third-party providers such as Snopes or PolitiFact. The screens hot in Figure~\ref{fig:facebook-disputed} is an example of this feature in action. In addition to adding this label, Facebook warns users with an alert box before they try to share a disputed story, but they are still allowed to share \cite{fb:2016}.

It remains to be seen how this new feature will affect fake news spreading. Again, access to Facebook data that could allow independent researchers to evaluate the effectiveness of such interventions. Without it, our understanding of changes in human behavior correlated with or caused by changes in the socio-technical platforms they inhabit, will be limited. This is a reason of concern for our research communities.

\section{Discussion}
% {\bf Eni, I leave this section to you, since you will need to decide what to include and conclude. I hope you have enough material to pick whatever you want to include in a short paper. However it is clear to me now that the whole topic needs a much longer journal article length treatment.}

What is the moral of the story? In the past, researchers were the ones discovering and documenting the misuse and abuse of socio-technical platforms by the hands of dubious actors with dubious agendas. \cite{mustafaraj:websci10} is only one such example. However, our discovery was possible only because we were collecting data in real-time, after having noticed some evidence of foul play. When we think about Twitter's approach to combating spammers, it seems reasonable that tweets created by misinformation-spreading accounts are automatically deleted and retracted from the entire network, once the accounts are suspended. However, the downside of such an approach is that it makes it impossible for researchers and fact-checkers to go back in time and study the origin of misinformation campaigns and the mechanism for spreading them. That is a severe limitation to research. The problem becomes even more pronounced in the content of fake news spreading on Facebook. 
Most Facebook groups are private and if they are the source for starting certain cascades of fake news, outside researchers cannot observe them in their early stages, missing crucial information that would lead to their understanding.     

Thus, it is not surprising that in the current situation created by the fake news plague, researchers didn't play a leading role in their discovery. It were journalists and not researchers in academia or Facebook and Google who raised concerns, but were not heard. This is worrisome. Facebook, by replacing humans with algorithms, has played a crucial role in fueling the fake news spreading phenomenon. Similarly, the ease with which Google enables earning ad money for page impression provided the financial incentives for the creation of the fake news industry.

In light of what we know so far, here is our open question to the research communities interested in information retrieval: 
\begin{quote}
in the context of the now omnipresent, web-based, socio-technical systems such as Facebook, Google, and Twitter, what tasks should be performed by humans and what tasks by algorithms? 
\end{quote}

Our communities should lead the way in providing answers to this question.

%ACKNOWLEDGMENTS are optional
%\section{Acknowledgments}
%Something 

\bibliographystyle{abbrv}
%\bibliography{websci17ref}  % sigproc.bib is the name of the Bibliography in this case

\end{document}